\documentclass[11pt]{article}
\usepackage[margin=1in]{geometry}
\usepackage{graphicx}
\usepackage{xspace}
\usepackage{amsmath}
\usepackage{placeins}

\def\beq{\begin{equation}}
\def\eeq#1{\label{#1}\end{equation}}
\def\eeqn{\end{equation}}

\def\beqa{\begin{eqnarray}}
\def\eeqa#1{\label{#1}\end{eqnarray}}
\def\eeqan{\end{eqnarray}}

\let\bar=\overbar

\def\Dslash{\not{\hbox{\kern-4pt $D$}}}
\def\dslash{\not{\hbox{\kern-2pt $\del$}}}

\def\msb{{\bar{\ssstyle M \kern -1pt S}}}

\def\Title#1{\begin{center} {\Large {\bf #1} } \end{center}}
\def\Author#1{\begin{center} {\normalsize {\sc #1} } \end{center}}
\def\Institution#1{\begin{center} {\normalsize {\it #1} } \end{center}}
\def\Abstract#1{\noindent {\normalsize {\bf Abstract:} {\normalfont #1}}}
\def\Conference{\vspace{4mm}\begin{raggedright} {\normalsize {\it Talk presented at the 2019 Meeting of the Division of Particles and Fields of the American Physical Society (DPF2019), July 29--August 2, 2019, Northeastern University, Boston, C1907293.} } \end{raggedright}\vspace{4mm}}

\begin{document}

%
%

\newcommand{\PZ}{\ensuremath{\text{Z}}\xspace}
\newcommand{\PW}{\ensuremath{\text{W}}\xspace}
\newcommand{\PH}{\ensuremath{\text{H}}\xspace}
\newcommand{\PV}{\ensuremath{\text{V}}\xspace}
\newcommand{\PX}{\ensuremath{\text{X}}\xspace}
\newcommand{\PAQb}{\ensuremath{\mathrm{\overline{b}}}}
\newcommand{\PQb}{\ensuremath{\mathrm{b}}}
\newcommand{\Pgt}{\ensuremath{\mathrm{\tau}}}
 
\newcommand{\PAQ}{\ensuremath{\mathrm{\overline{q}}}}
\newcommand{\PQ}{\ensuremath{\mathrm{q}}}

\newcommand{\cPqt}{\ensuremath{\mathrm{t}}}
\newcommand{\bbbar}{\PQb{}\PAQb\xspace}
\newcommand{\ttbar}{\ensuremath{{\mathrm{t}\overline{\mathrm{t}}}}\xspace} 

\newcommand{\qqbar}{\PQ{}\PAQ\xspace}

\newcommand{\mtn}{\ensuremath{m_{\cPqt}}\xspace}
\newcommand{\mwn}{\ensuremath{m_{\PW}}\xspace}
\newcommand{\losttwn}{\ensuremath{\text{lost}~\cPqt/\PW}\xspace}
\newcommand{\qgn}{\ensuremath{\text{q/g}}\xspace}
\newcommand{\TeV}{\ensuremath{\,\text{Te\hspace{-.08em}V}}\xspace}
\newcommand{\GeV}{\ensuremath{\,\text{Ge\hspace{-.08em}V}}\xspace}
\newcommand{\fbinv} {\mbox{\ensuremath{\,\text{fb}^\text{$-$1}}}\xspace}

\newcommand{\mtbkg}{\ensuremath{\mtn~\text{background}}\xspace}
\newcommand{\mwbkg}{\ensuremath{\mwn~\text{background}}\xspace}
\newcommand{\losttwbkg}{\ensuremath{\losttwn~\text{background}}\xspace}
\newcommand{\qgbkg}{\ensuremath{\qgn~\text{background}}\xspace}

\newcommand{\wjets}{\ensuremath{\PW+~\text{jets}}\xspace}

\newcommand{\hh}{\ensuremath{\PH\PH}\xspace}
\newcommand{\mbb}{\ensuremath{m_{\bbbar}}\xspace}
\newcommand{\mhh}{\ensuremath{m_{\hh}}\xspace}
\newcommand{\mx}{\ensuremath{m_{\PX}}\xspace}

\newcommand{\bbbb}{\ensuremath{\PQb{}\PAQb{}\PQb{}\PAQb}\xspace}
\newcommand{\bbtautau}{\ensuremath{\PQb{}\PAQb{}\Pgt{}\Pgt}\xspace}
\newcommand{\bbWW}{\ensuremath{\PQb{}\PAQb{}\PW{}\PW^*}\xspace} 

\newcommand{\hbb}{\ensuremath{\PH\rightarrow\bbbar}\xspace} 
\newcommand{\wqq}{\ensuremath{\PW\rightarrow\qqbar'}\xspace} 

\newcommand{\pt}{\ensuremath{p_\mathrm{T}}\xspace} 

\Title{Searching for resonant $\PH\PH$ production in the $\bbbar\qqbar'\ell\nu$ final state at CMS}

\Author{Nickolas McColl on behalf of the CMS Collaboration}
\Institution{University of California, Los Angeles \\ E-mail: nick.mccoll@cern.ch}


\Abstract{New, massive bosons could be found with the LHC. Theories with warped extra dimensions and supersymmetry predict the existence of such resonances, which for some model parameters, have a significant branching fraction to two Higgs bosons. A search for such particles in the $\PX\rightarrow\PH\PH\rightarrow\bbbar\qqbar'\ell\nu$ channel with the CMS detector is presented. The analysis uses data collected during Run 2 of the LHC at a centre-of-mass energy of 13 TeV. Background is suppressed by reconstructing the full $\PH\PH$ decay chain using jet substructure techniques and the identification of leptons with nearby, boosted jets. A two-dimensional template fit in the plane of resonance the mass and the $\PH\rightarrow\bbbar$ mass is used to characterize potential signal with this final state.}

\Conference

\section{Introduction}

The Higgs boson ($\PH$) is an important tool in the search for new physics at the CERN LHC. While its pair production is rare in the standard model (SM), there exists a broad range of theories that predict new bosons that decay to $\PH\PH$. These include supersymmetry~\cite{Nilles:1983ge} and Randall-Sundrum models of warped extra dimensions~\cite{Randall:1999ee}, for which the new bosons would be spin-0 radions or spin-2 bulk gravitons.

A search for such new particles $\PX$ in the $\PX\rightarrow\PH\PH\rightarrow\bbWW$~\cite{Sirunyan:2019quj} decay channel is presented here. It is performed on a data set collected by the CMS detector~\cite{Chatrchyan:2008zzk} corresponding to an integrated luminosity of $35.9\fbinv$ of proton-proton collisions at $\sqrt{s}=13 \TeV$. The potentially large SM background is reduced by analyzing events in which one $\PW$ boson decays to hadrons and the the 
other decays to an electron or muon and a neutrino. The search is optimized for particle mass $\mx>0.8$.  Such large values of \mx lead to the distinctive experimental signature of two back-to-back, collimated Higgs boson decays.

\section{Event selection}
Events are only included in the search if all final state particles from an $\PH\PH\rightarrow\bbWW$ decay can be associated to reconstructed objects. The charged lepton is identified as an electron or muon that passes an isolation criteria optimized for boosted particle decays. The neutrino transverse momentum is reconstructed as missing transverse momentum, while its longitudinal momentum is obtained by imposing a Higgs boson mass constraint on the reconstructed $\PH\rightarrow\PW\PW^*$ decay.

\begin{figure}[ht!]
\centering
\includegraphics[width=0.45\textwidth]{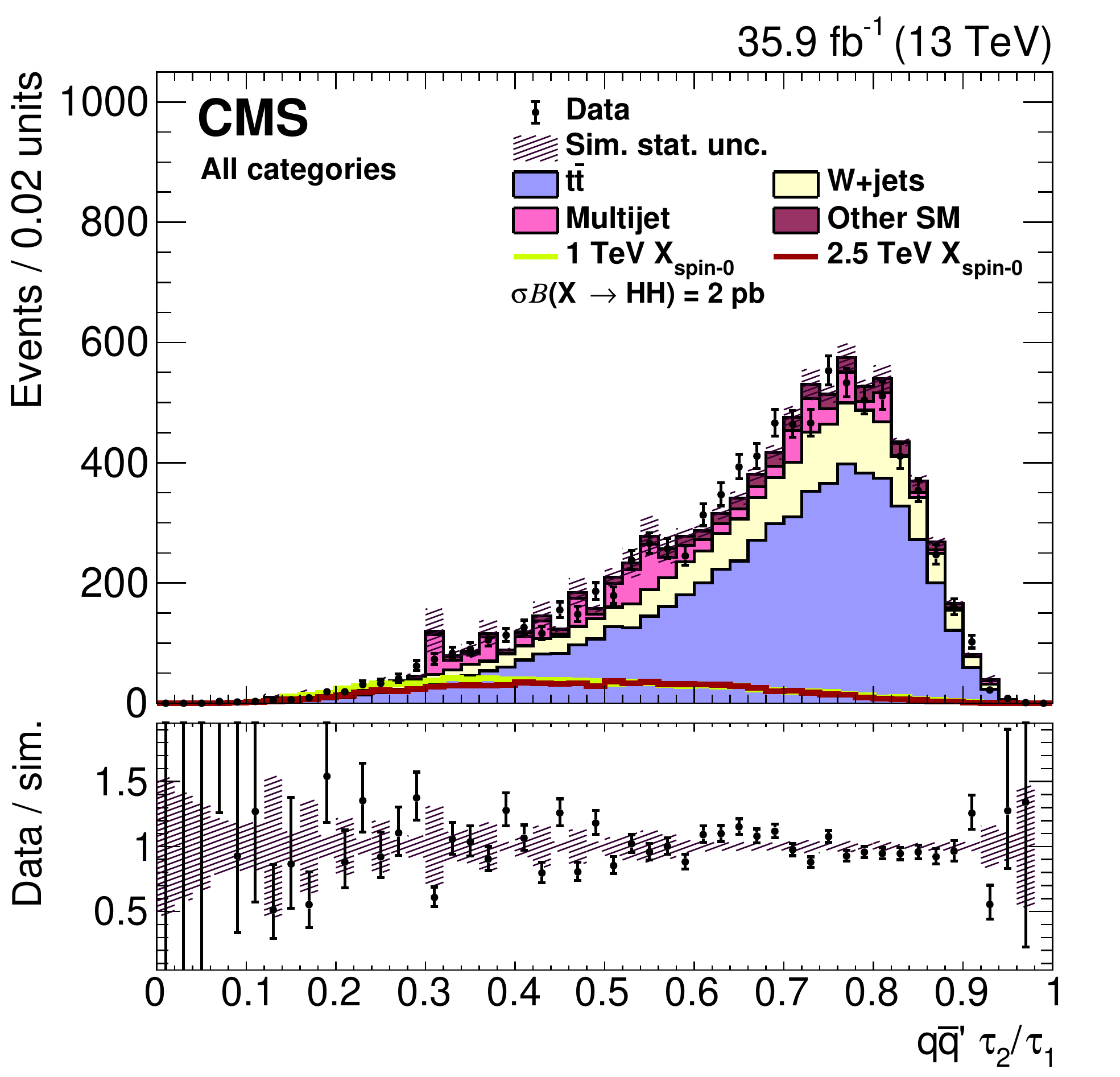}
\includegraphics[width=0.45\textwidth]{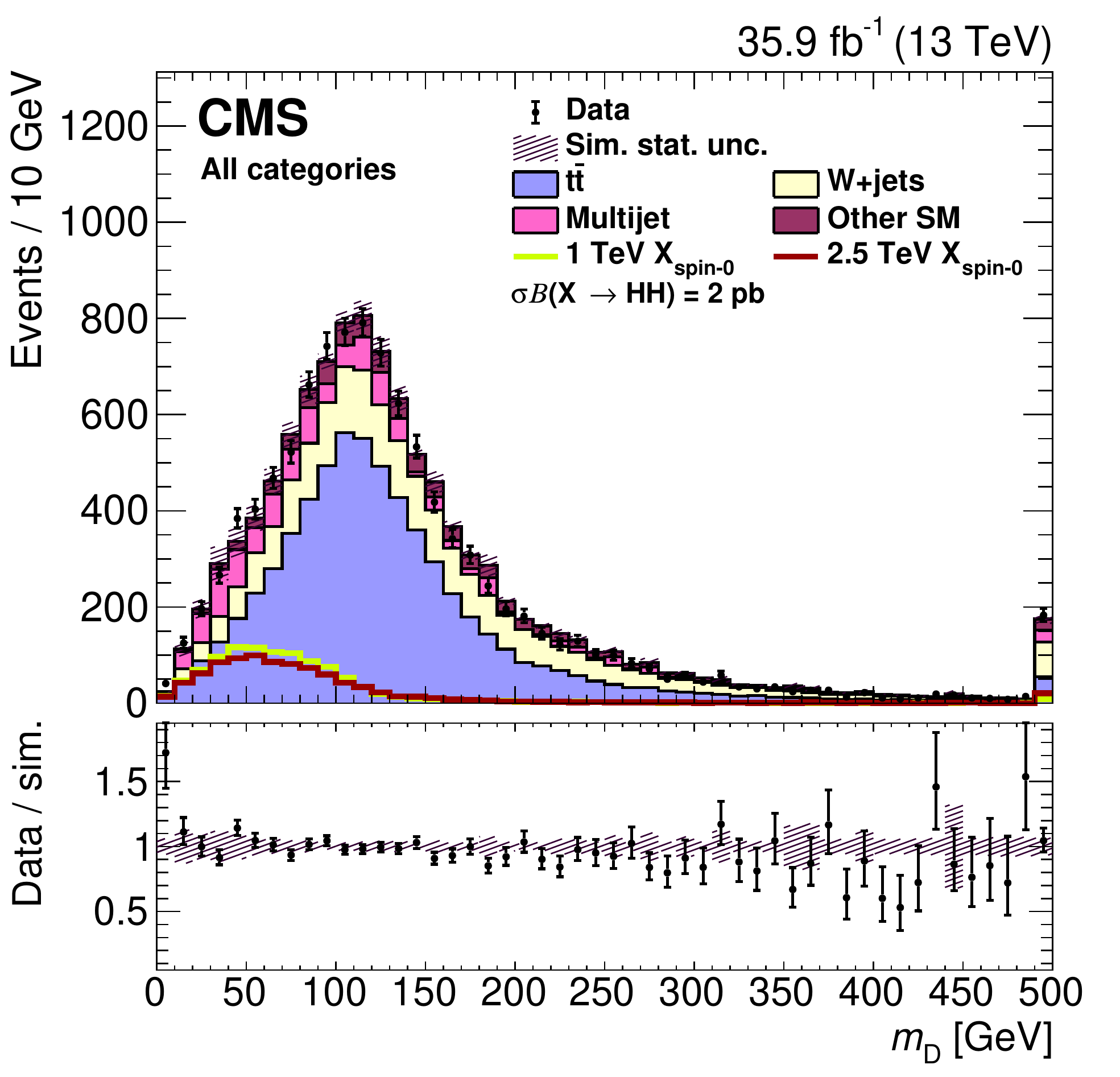}

\caption{Variables used to suppress background events. Left: $\tau_2/\tau_1$ of the merged \wqq decay. Right: $m_{\mathrm{D}}$, which is the $\pt \Delta R /2$ of the reconstructed $\PH\rightarrow\PW\PW^*$ decay~\cite{Sirunyan:2019quj}.}
\label{fig:eventsel}
\end{figure}

The hadronic $\PW$ boson decay is reconstructed as a single large-radius jet near the lepton. This jet is clustered according to the anti-$k_\mathrm{T}$ algorithm with $R=0.8$~\cite{Cacciari:2008gp,Cacciari:2011ma} and is further required to have substructure consistent with a two-prong decay. First, the ``modified mass drop tagger'' algorithm~\cite{Dasgupta:2013ihk,Butterworth:2008iy}, also known as the ``soft drop'' (SD) algorithm, with angular exponent $\beta = 0$, soft cutoff threshold $z_{\mathrm{cut}} < 0.1$, and characteristic radius $R_{0} = 0.8$~\cite{Larkoski:2014wba}, is applied to this jet to identify two subjets. Second, the ``N-subjettiness'' ratio $\tau_2/\tau_1$~\cite{Thaler:2010tr} of this jet is required to be $<0.75$. The $\tau_2/\tau_1$ distribution is shown in Fig.~\ref{fig:eventsel}.

The $\PH\rightarrow\bbbar$ decay is also reconstructed as a single large-radius jet, but in this case the jet and lepton are required to have large angular separation ($\Delta\phi>2$). The jet is required to have two SD subjets and their invariant mass \mbb is a search variable. This jet is also identified as likely originating from two b hadron decays by requiring that the individual subjets are b tagged.

Two signal characteristics are then used to further suppress background. The first is that jets in signal events are produced more centrally in $|\eta|$ than those in QCD multijet and $\wjets$ events. These high $|\eta|$ background events are removed by requiring that the $\pt$ of the reconstructed $\PH\rightarrow\PW\PW^*$ decay divided by the reconstructed resonance mass $\mhh$ is $>0.3$. The second characteristic is that the reconstructed $\PW$ bosons in signal events are more collimated than those in background events. These wide-angle events are emoved by requiring $\pt \Delta R /2 < 125\GeV$. Here the $\pt$ is that of the reconstructed $\PH\rightarrow\PW\PW^*$ decay and the angular distance is between the two $\PW$ bosons. This distribution of this variable is shown in Fig.~\ref{fig:eventsel}.

\section{Signal extraction}

\begin{figure}[ht!]
\centering
\includegraphics[width=0.45\textwidth]{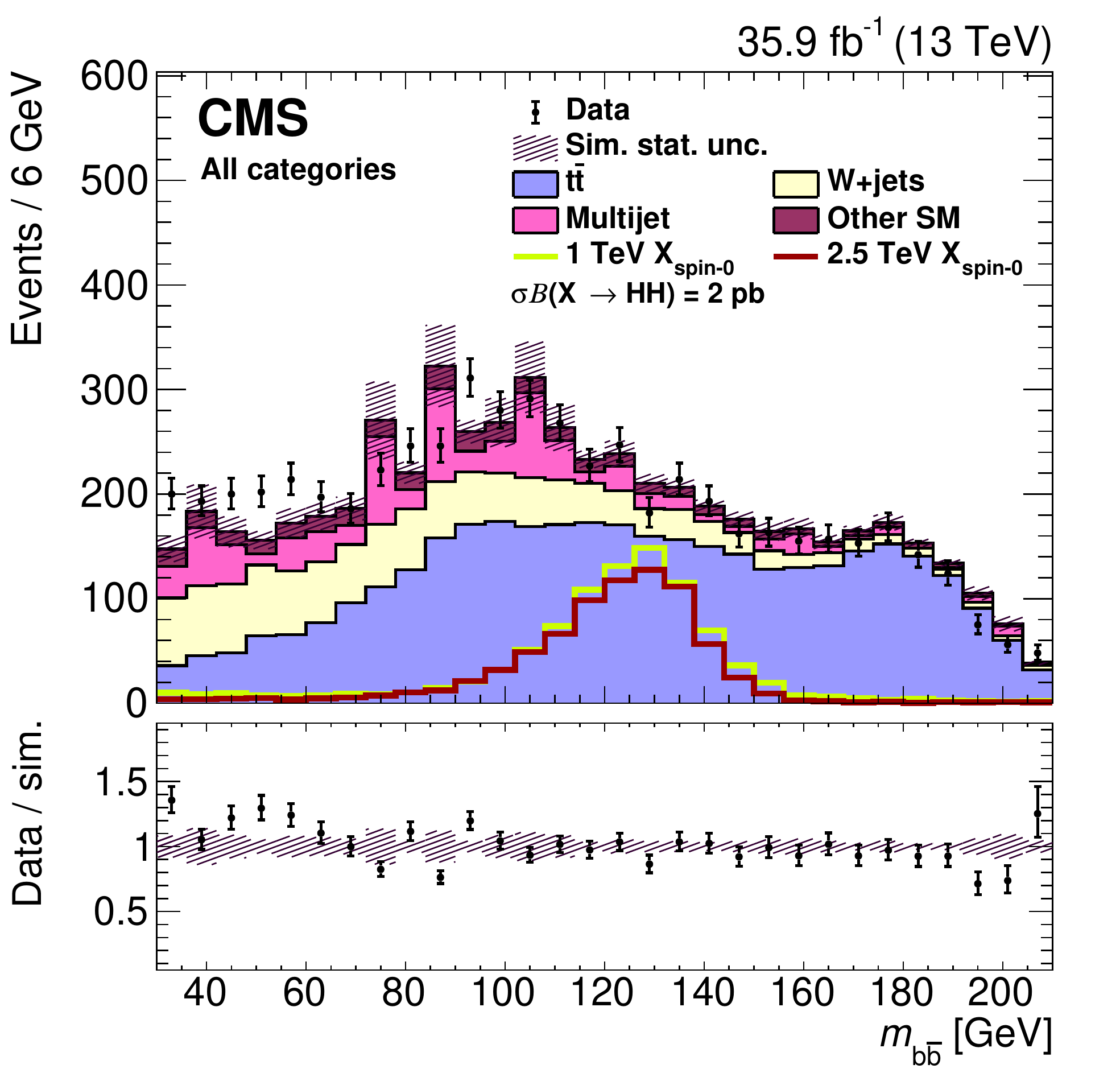}
\includegraphics[width=0.45\textwidth]{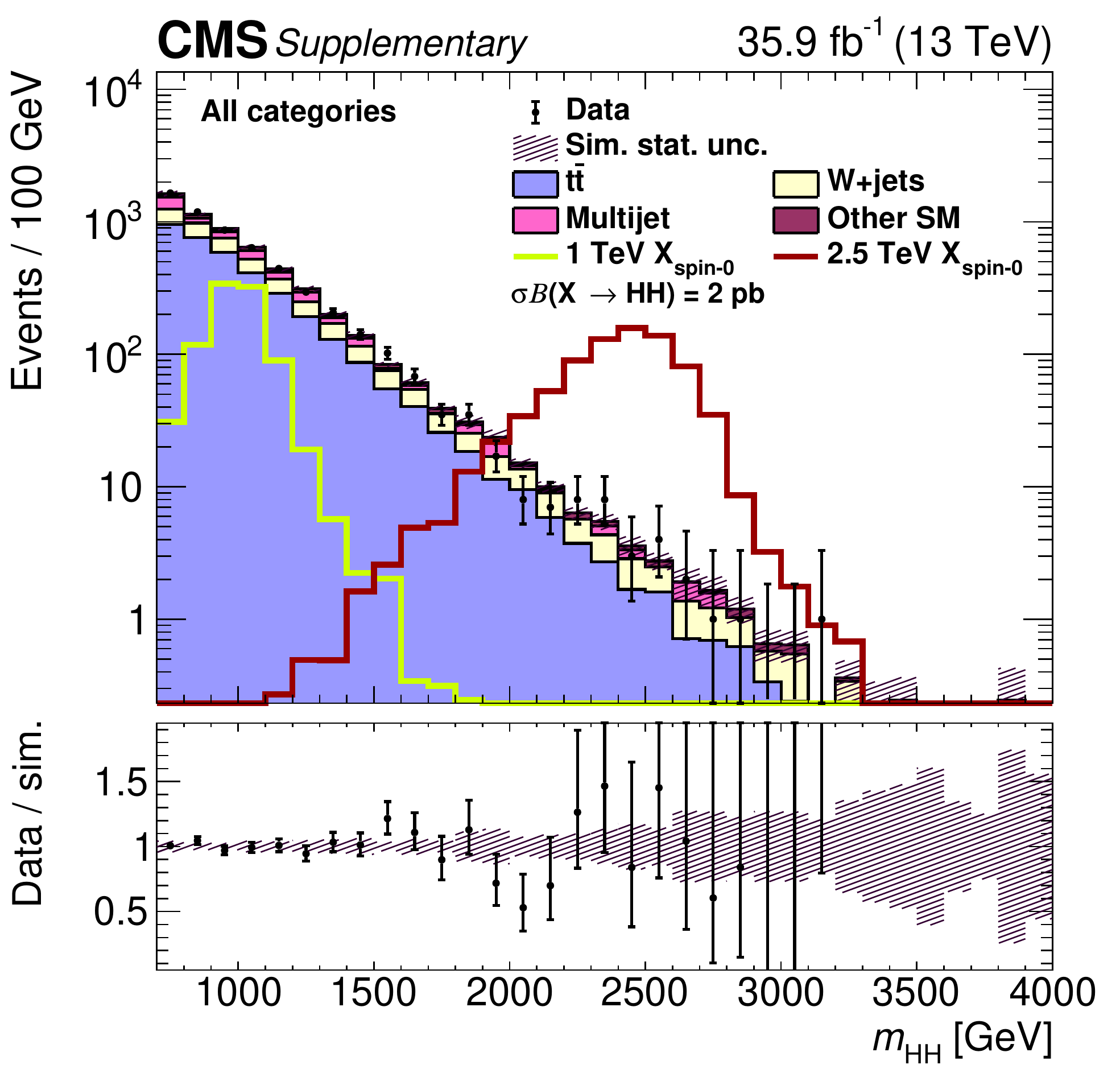}

\caption{The two discriminating variables used to characterize signal: \mbb (left) and \mhh (right)~\cite{Sirunyan:2019quj}. }
\label{fig:masses}
\end{figure}

The distinguishing characteristic of signal is a peak in the two-dimensional plane of \mbb and \mhh. There is no such peak for background, which is primarily composed of \ttbar events. This leads to the characterization of the analysis as identifying a two-dimensional peak over a smooth background.  The \mbb and \mhh distributions are shown in Fig.~\ref{fig:masses}.

The background and signal yields are simultaneously estimated using a maximum likelihood fit to the two two-dimensional mass plane. Templates are used to model signal and background, which are derived from simulation. Differences between data and simulation (systematic uncertainties) are included in the fit as nuisance parameters.

The background model is validated by estimating the SM background in two background-enhanced control regions. The first is used to test the modelling of \ttbar by requiring that events have an extra b-tagged jet. The second is used to test the modeling of the QCD multijet and \wjets processes by requiring that no subjets in the $\PH\rightarrow\bbbar$ jet are b tagged. In both cases the background is found to be modelled well.

\section{Results and conclusion}

No significant excesses are observed. The quality of the fit to the search region data is quantified with a goodness-of-fit test. A generalization of the $\chi^2$ test of Poisson statistics is used~\cite{Baker:1983tu}. The p-value of the test is 0.6, indicating tha the model fits the data well.

\begin{figure}[ht!]
\centering
\includegraphics[width=0.45\textwidth]{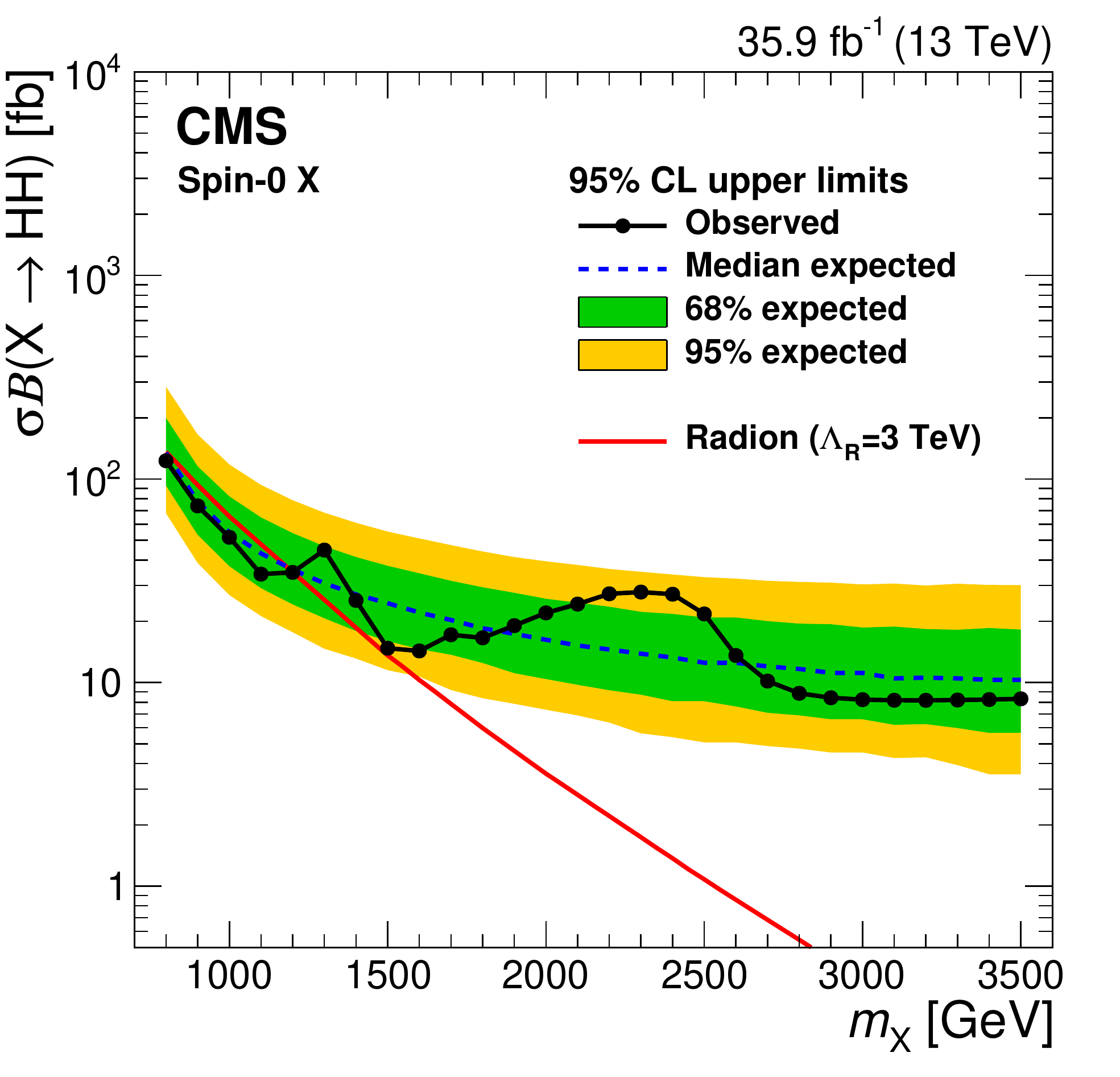}
\includegraphics[width=0.45\textwidth]{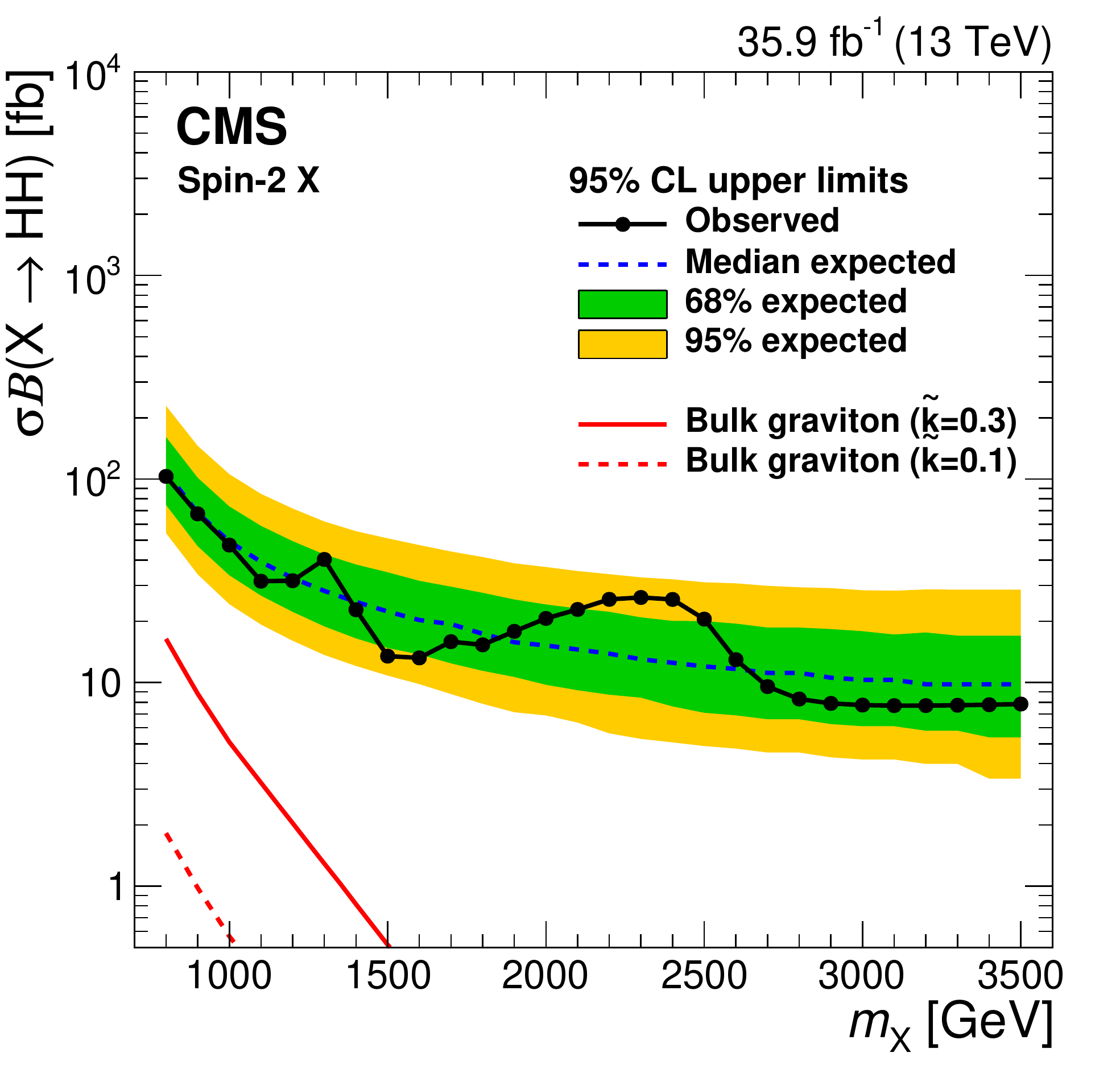}
\caption{The 95\% confidence level upper limits for spin-0 (left) and spin-2 (right) bosons as a function of \mx~\cite{Sirunyan:2019quj}.}
\label{fig:bbWWLimits}
\end{figure}

The results are interpreted as upper limits on the product of the $\PX$ production cross section and the $\PX\to\PH\PH$ branching fraction. The 95\% confidence level upper limits is shown in Fig.~\ref{fig:bbWWLimits}. These are the best results to date for resonances decaying to this final state. It has similar sensitivity to searches in other decay channels for $\mx<1.5\TeV$, bringing new sensitivity to resonant $\PH\PH$ production.

The result presented here is only one piece of the rich CMS $\PH\PH$ program. It is comprised of a diverse set of analyses, each targeting a different mass scale or decay channel. No hints of new physics have been observed so far, but only a fraction of the full LHC Run 2 data set has been analyzed. Future results that exploit the full Run 2 data set are expected to be much more sensitive to this important signal.

\FloatBarrier

\bibliographystyle{JHEP}
\bibliography{sources.bib}
 
\end{document}